\pdfoutput=1
\documentclass[journal]{IEEEtran}
\usepackage{cite}
\usepackage{amsfonts}

\usepackage{amssymb,latexsym}
\usepackage{multirow}
\usepackage[pdftex]{graphicx}
\DeclareGraphicsExtensions{.pdf,.png}
\usepackage{epstopdf}
\usepackage{float}
\usepackage{color}
\usepackage{enumerate}
\usepackage{enumitem}
\usepackage{booktabs}
\usepackage{tabularx}
\usepackage{amsmath}
\usepackage{algorithmic}
\usepackage{array}
\usepackage{color}
\usepackage{fixltx2e}
\usepackage{hyperref}
\usepackage{url}
\allowdisplaybreaks[4]
\usepackage{subfigure}
\usepackage[numbers]{natbib}
\def \eg{\emph{e.g.}} 
\begin{document}
\title{DualVGR: A Dual-Visual Graph Reasoning Unit for Video Question Answering}
%
\author{Jianyu~Wang,
        Bing-Kun~Bao\IEEEauthorrefmark{1}, Changsheng~Xu,~\IEEEmembership{Fellow,~IEEE}\thanks{This work is supported by the National Key Research \& Development Plan of China  2020AAA0106200, the National Natural Science Foundation of China under Grant 61936005, 62036012, 61872424, the Natural Science Foundation of Jiangsu Province(Grants No BK20200037).}\thanks{Jianyu Wang, Bing-Kun Bao are with the College of Telecommunications \& Information Engineering, Nanjing University of Posts and Telecommunications, Nanjing, 210003, China (e-mail: jianyuwang.work@outlook.com; bingkunbao@njupt.edu.cn).}\thanks{Changsheng Xu is with the National Laboratory of Pattern Recognition, Institute of Automation, Chinese Academy of Sciences, Beijing 100190, China, and also with the School of Artificial Intelligence, University of Chinese Academy of Sciences, Beijing 100049, China (e-mail: csxu@nlpr.ia.ac.cn).}\thanks{\IEEEauthorrefmark{1} Corresponding author. E-mail: bingkunbao@njupt.edu.cn}.}


%
%

\markboth{IEEE TRANSACTIONS ON MULTIMEDIA,~Vol.~14, No.~8, August~2021}%
{Shell \MakeLowercase{\textit{et al.}}: Bare Demo of IEEEtran.cls for IEEE Journals}
%



\maketitle
\begin{abstract}
Video question answering is a challenging task, which requires agents to be able to understand rich video contents and perform spatial-temporal reasoning. However, existing graph-based methods fail to perform multi-step reasoning well, neglecting two properties of VideoQA: (1) Even for the same video, different questions may require different amount of video clips or objects to infer the answer with relational reasoning; (2) During reasoning, appearance and motion features have complicated interdependence which are correlated and complementary to each other. Based on these observations, we propose a Dual-Visual Graph Reasoning Unit (DualVGR) which reasons over videos in an end-to-end fashion. The first contribution of our DualVGR is the design of an explainable Query Punishment Module, which can filter out irrelevant visual features through mutiple cycles of reasoning. The second contribution is the proposed Video-based Multi-view Graph Attention Network, which captures the relations between appearance and motion features. Our DualVGR network achieves state-of-the-art performance on the benchmark MSVD-QA and SVQA datasets, and demonstrates competitive results on benchmark MSRVTT-QA datasets. Our code is available at \href{https://github.com/MM-IR/DualVGR-VideoQA}{https://github.com/MM-IR/DualVGR-VideoQA}.
\end{abstract}

%
\IEEEpeerreviewmaketitle

\section{Introduction}
VideoQA is a challenging and high-level multimedia task \cite{zhu2020multimedia}, which requires the agents to understand videos and perform relational reasoning according to questions based on visual, textual, as well as spatial-temporal contents. Since the input of VideoQA is a sequence of frames, there are two differences between ImageQA and VideoQA: (1) In addition to appearance information, VideoQA also needs to understand the motion information to answer the questions. (2) VideoQA requires to perform spatio-temporal reasoning over the objects, while ImageQA only requires spatial reasoning over the objects. Therefore, scene graphs \cite{teney2017graph, li2019relation, hu2019language} and neural-symbolic reasoning frameworks \cite{hu2017learning,yi2018neural,shi2019explainable,li2019perceptual}, which are used in ImageQA, are hard to be implemented in VideoQA as the agents have to address the issues of comprehensive representation (\eg{ appearance and motion information}) and multi-step reasoning. 

In order to solve the above challenges, this work focuses on performing multi-step reasoning via Graph Networks for VideoQA. Previous reasoning-based methods can be divided into four categories based on their frameworks. The first group \cite{li2019beyond, zhao2020open, zhao2019multi, zhao2019long} implements spatial and temporal attention mechanism to iteratively select useful information to answer the questions. The second group \cite{xu2017video, gao2018motion} focuses on memory-based network, which is quite popular in TextQA. However, these methods neglect the visual relation information when performing multi-step reasoning. The third group \cite{le2020hierarchical} aims to perform relational reasoning via a simple module, like relation network. However, this module can only model a limited number of objects. The fourth group \cite{huang2020location, jiang2020reasoning} aims to use graph neural networks to integrate relation information into their frameworks by considering GNN's powerful representation ability on relation modeling. GNN is a novel relation encoder that captures the inter-object relations beyond static object/region detection, thus enabling reasoning with rich relational information. Compared with relation network, graph neural network is more flexible and powerful in relational reasoning, thus we follow this group in our work.  
\begin{figure}[t]
    \centering
    \includegraphics[scale=0.14]{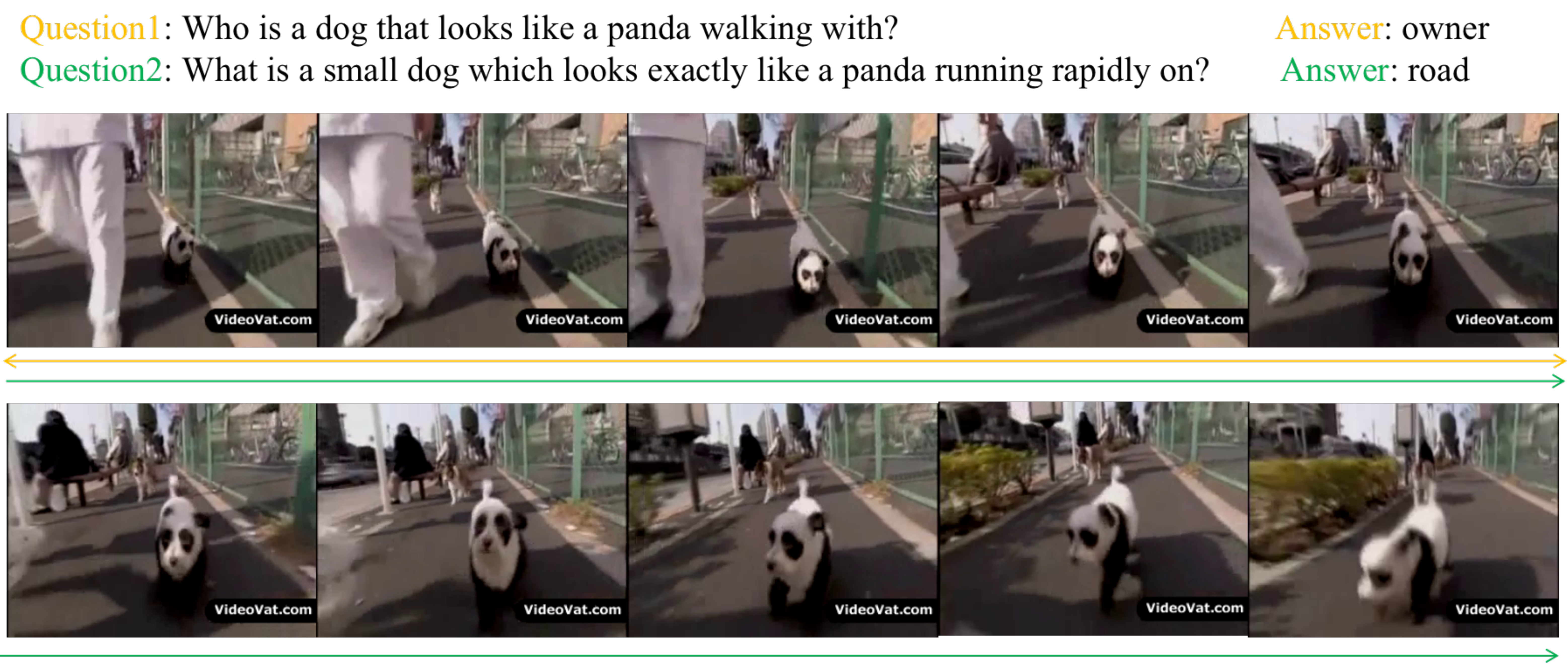}
    \caption{Answering questions requires agents to utilize both appearance and motion information, and usually requires them to be capable of multiple steps of spatio-temporal reasoning \cite{xu2016msr}. (a) \textit{Question1}: Not all video shots or objects are necessary to answer the question. (b) \textit{Question2}: In each reasoning cycle, appearance and motion features are usually associated and complementary to each other.}
    \label{fig:Example}
\end{figure}

However, existing graph-based methods \cite{huang2020location, jiang2020reasoning} neglect two key attributes of VideoQA task when performing reasoning: (1) Not all video shots or objects are correlated to the question. As illustrated in Fig. \ref{fig:Example}., to answer \textit{Question1: Who is a dog that looks like a panda walking with?}, only question-related video clips are needed to infer the answer. (2) Appearance and motion features are associated and complementary to each other in each reasoning step. Using Fig. \ref{fig:Example} as an example again, when answering \textit{Question2: What is a small dog which looks exactly like a panda running rapidly on?}, it is necessary for agents to utilize both appearance and motion information. Specifically, in order to understand \textit{a small dog which looks exactly like a panda}, agents need to rely on appearance information to find the dog. Then, they have to analyze the motion information to infer the action of the dog: \textit{running rapidly on the road}. In short, The appearance information could provide clues to pay attention to certain motion information to answer the questions, vice versa.

Motivated by these observations, we devise a novel graph-based reasoning unit named Dual-Visual Graph Reasoning Unit (DualVGR), which is stacked in an iterative manner to perform multi-step reasoning. At first, we design \textit{Query Punishment Module} to generate query-guided masks to allow a limited number of question-related video features into relational reasoning. Then, in order to fully capture the multi-view visual relation information, we propose \textit{Video-based Multi-view Graph Attention Network} to reveal the relation between appearance and motion channels. The proposed graph network includes two graphs for each visual channel. The first graph aims to learn the underlying complementary relation within each specific visual space, and the other is to learn the concomitant and correlated relation between appearance and motion features. The losses of video-based multi-view graph network are consistency constraint loss and disparity constraint loss. Consistency constraint loss is to enhance the commonality between appearance and motion features, while disparity constraint loss is to enhance the heterogeneity between them.   

The main contributions of this work can be summarized as follows. (1) We propose a DualVGR unit, a multimodal reasoning unit enabling the represention of rich interactions between question and video clips. This unit includes two components. One is an explainable Query Punishment Module to filter out irrelevant visual-based information. It has been demonstrated good results in both short and long compositional questions during multiple cycles of reasoning. The other is a Video-based Multi-view Graph Attention Network to perform spatial-temporal relational reasoning such that the relations between appearance and motion can be adaptively captured; (2) We incorporate our DualVGR unit into a full DualVGR network by stacking it in an iterative manner. Through multi-step reasoning, our method achieves state-of-the-art or competitive results on several mainstream datasets: MSVD-QA \cite{xu2017video}, MSRVTT-QA \cite{xu2016msr} and SVQA \cite{song2018explore}.
\section{Related Work}
In this section, we review the recent studies related to visual question answering, which can be divided into two sub-categories: image question answering and video question answering.
\subsection{Image Question Answering}
There are three research directions about image question answering. The first line, namely monolithic method \cite{li2019perceptual, nie2012beyond}, locates the most relevant visual region of images based on attention mechanism, then projects visual features and textual features together into a common latent space through a single step. Antol \textit{et al.} \cite{antol2015vqa} combine the visual features and the textual features via multimodal pooling, such as addition and concatenation, then map them into a unified space. However, multimodal pooling methods do not well capture the complex associations between two modalities due to their different distributions. Therefore, some approaches \cite{tenenbaum2000separating} related to Bilinear Pooling, which has been used to integrate different CNN features for fine-grained image recognition \cite{lin2015bilinear}, have been proposed. However, Bilinear Pooling needs a huge number of parameters and the dimensionality of output feature is usually too high. Some extended versions are proposed to handle these issues, such as Multimodal Compact Bilinear Pooling (MCB) \cite{fukui2016multimodal}, Hadamard Product for Low-rank Bilinear Pooling (MLB) \cite{kim2016hadamard}, Multimodal Factorized Bilinear Pooling (MFB) \cite{yu2017multi} and Multimodal Factorized High-order pooling (MFH) \cite{yu2018beyond}. Monolithic methods have been demonstrated useful in some cases, but failed to perform well in long and compositional questions.

The second line focuses on modeling the multi-step interaction process through a recurrent cell \cite{yang2016stacked, nguyen2018improved, yu2019deep, xiong2016dynamic,nie2019multimodal}. Nguyen and Okatani \cite{yang2016stacked}, Yang \textit{et al.} \cite{nguyen2018improved} stack the attention layers to model the multi-step interaction process. Xiong \textit{et al.} \cite{xiong2016dynamic} introduce a novel dynamic memory network to iteratively retrieve meaningful visual contents. Meanwhile, some approaches \cite{teney2017graph, li2019relation, hu2019language} aim to integrate relational reasoning in each step, especially graph-based methods. Li \textit{et al.} \cite{li2019relation} model multi-type inter-object relations via a graph attention mechanism to learn question-adaptive relation representations.

The third line is neural-symbolic reasoning. This kind of methods decompose the whole task into some subtasks, and design several Neural Module Networks (NMN) \cite{andreas2016neural,hu2017learning,yi2018neural,gan2017vqs,mao2018neuro,yi2019clevrer,chen2021grounding} to solve those subtasks. Andreas \textit{et al.} \cite{andreas2016neural} firstly utilize an off-the-shelf parser to decompose the compositional question into logical expressions, then construct a monolithic network for each subtask. Hu \textit{et al.} \cite{hu2017learning} claim that a brittle off-the-shelf semantic parser would lead to bad performance, hence they design a seq-to-seq RNN to end-to-end predict instance-specific layout. These kinds of methods receive good performances on synthetic datasets \cite{johnson2017clevr}, as the questions are easy to parse into subtasks. Besides, Yi \textit{et al.} \cite{yi2019clevrer} propose a CLEVRER dataset for exploring the problem of temporal and causal reasoning in videos. This dataset further promotes the development of the neuro-symbolic reasoning research in video-related tasks.
\begin{figure*}[ht]
    \centering
    \includegraphics[scale=0.32]{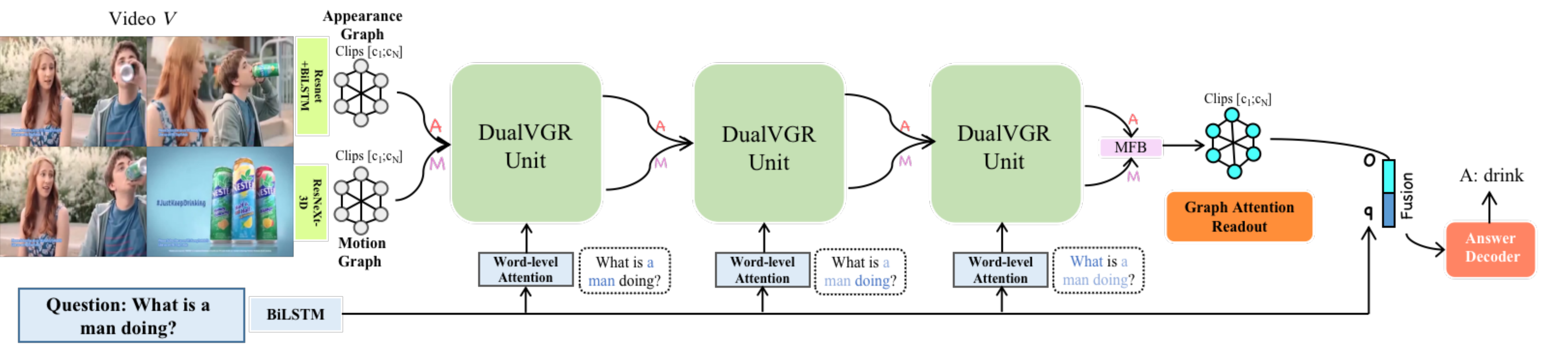}
    \caption{DualVGR architecture for VideoQA. First, we divide the whole video into certain clips, and extract the appearance and motion features from each clip. Meanwhile, Question is embed with BiLSTM. Then, with these representations, we send them to our stacked DualVGR unit to iteratively model rich interactions between video clips. Next, we fuse appearance and motion features of each clip by Bilinear Fusion, and utilize attention mechanism to fuse features of all clips to obtain the final visual vector. Finally, we concatenate the visual vector with our question vector to predict the answers.}
    \label{fig:Framework}
\end{figure*}
\subsection{Video Question Answering}
Different from ImageQA, VideoQA is more challenging as videos contain much more complex patterns than a single image. To tackle this problem, agents have to fully comprehend the temporal structure of videos. Jang \textit{et al.} \cite{jang2017tgif} encode appearance feature and motion feature with ResNet \cite{he2016identity} and C3D \cite{ji20123d} respectively, then design a spatial and temporal attention mechanism to select different regions during multi-step reasoning. Later, with the success of Dynamic Memory Network \cite{xiong2016dynamic} in ImageQA, some work \cite{xu2017video, gao2018motion} applies memory-based network to VideoQA tasks. Xu \textit{et al.} \cite{xu2017video} propose an Attention Memory Unit (AMU), which gradually refines its attention over the appearance and motion features by using question as guidance. Considering that motion and appearance features are correlated in the reasoning process, Gao \textit{et al.} \cite{gao2018motion} propose a motion-appearance co-memory network and use a temporal convolutional and deconvolutional neural network to generate multi-level contextual facts. However, their methods do not consider the relations between different objects.

Relation information is obviously an important clue in the reasoning process. Le \textit{et al.} \cite{le2020hierarchical} propose a Conditional Relation Network (CRN) to model the relations between visual objects, and stack CRNs to perform multi-step relational reasoning. However, the limitation of relation network-based methods is that they can only process a limited number of objects at a time. Therefore, graph neural network comes into researchers' sight. Huang \textit{et al.} \cite{huang2020location} point out that previous work neglects the interactions among objects in each frame and design a location-aware graph convolutional network to learn a relation-enhanced visual feature encoder. Jiang \textit{et al.} \cite{jiang2020reasoning} propose an undirected heterogeneous graph with each video snippet and question word as node to integrate correlations of both inter- and intra-modality in an uniform module. In these graph-based methods, all the objects, even those irrelevant to the question, are engaged into graph construction without discrimination, which could bring lots of uninformative noises into relational reasoning.  

\section{Our Approach}
The task of VideoQA can be described as follow. Given a video $V$ and the corresponding question $q$, agents aim to infer the answer $a$ correctly. Formally, the prediction \~{a} is given by classification scores:
\begin{equation}
    \Tilde{a} = \mathop{argmax}\limits_{a \in A}P_\theta\left(a | q,V\right),
\end{equation}
where $\theta$ is our trainable model parameters. 

The proposed DualVGR framework is depicted in Fig. \ref{fig:Framework}. Firstly, each video is divided into several clips, and we extract appearance and motion features for each clip. Meanwhile, the corresponding question is embed with BiLSTM. Secondly, appearance features, motion features and the corresponding question features are input into the our stacked DualVGR unit iteratively. We use each DualVGR unit to determine the attention within video guided by the question, and model the relational reasoning by stacking the DualVGR units. Thirdly, we fuse appearance and motion features of each clip by Multimodal Factorized Bilinear pooling (MFB) \cite{yu2017multi}, and utilize attention mechanism to fuse features of all clips to obtain the final visual vector. Finally, we concatenate the visual vector with our question vector to predict the answers.

\subsection{Visual and Linguistic Representation} 
We firstly divide the whole video $V$ of $L$ frames into $N$ equal length clips $C = \left(C_1,\dots,C_N\right)$. Therefore, each video clip $C_i$ contains $F$ frames, where $F = \lfloor{L/N}\rfloor$. We extract two sources of information from each video clip, which are appearance features and motion features. The appearance features of each video clip $C_i$ are denoted as $V^i_a = \{v^i_{a,1},v^i_{a,2},\dots,v^i_{a,F}\} \in \mathbb{R}^{2048 \times F}$, where the subscript $a$ indicates appearance. The motion features of each clip are represented as $V^i_m \in \mathbb{R}^{2048 \times 1}$, where the subscript $m$ indicates motion. For fair comparison with the state-of-the-art methods (in Section \ref{sec:experiments}) \cite{le2020hierarchical}, $V^i_a$ are the \textit{pool5} output of ResNet101 and $V^i_m$ are extracted from ResNeXt-101.

We follow the previous work \cite{le2020hierarchical,yang2019question} to process questions, which embed the words of each question into fixed-length vectors initialized with 300-dimension GloVe \cite{pennington2014glove} as $W = \{w_i: 1 \leqslant i \leqslant L_q, w_i \in \mathbb{R}^{300 \times 1}\}$, where $L_q$ is the length of each question. Then, we pass these word-embeddings into a BiLSTM Network to get context-aware embedding vectors. Different from the previous methods which just use the last hidden state of LSTM to represent the questions without deep analysis, we process all the hidden state sequences generated by BiLSTM $Q \in \mathbb{R}^{d \times L_q}$ in each step, which will be described latter in Section $B$. Furthermore, in order to further perform graph-based reasoning, appearance features should have the same dimension as motion features. Therefore, we further pass the appearance features of each video clip $V^i_a$ into a BiLSTM network, and project the motion features $V^i_m \in \mathbb{R}^{d \times 1}$ of each video clip $C_i$ into a fixed-size vector with dimension $d$. The generated video subclip-based appearance features $V^i_a \in \mathbb{R}^{d \times 1}$ and question features $Q \in \mathbb{R}^{d \times L_q}$ denote as:
\begin{equation}
\left\{
\begin{aligned}
   \{V^i_a, O\} & = BiLSTM\left(V^i_a, \theta^{\left(v\right)}_{BiLSTM}\right),\\
   \{E^Q, Q\} & = BiLSTM\left(W, \theta^{\left(q\right)}_{BiLSTM}\right),
\end{aligned}
\right.
\end{equation}
where $O \in \mathbb{R}^{d \times L_q}$ is the hidden state of appearance-based BiLSTM, and $E^Q \in \mathbb{R}^{d \times 1}$ is the output of the last hidden units and represents the global features of the question.
\subsection{The proposed DualVGR} 
To answer the question based on the given video, the designed agent needs to have the ability of locating on the question-related video clips, understanding the video with both appearance and motion channels, and performing multi-step reasoning. The proposed DualVGR Unit determines the question-related video clips, mines the relation within/between appearance and motion, and combines complementing relation features with corresponding visual features, then stacks the DualVGR units for multi-step relational reasoning. It consists of two modules, Query Punishment Module and Video-based Multi-view Graph Attention Network, as shown in Fig. 3. Query Punishment Module is designed to filter out the irrelevant video clips by employing query-guided mask on video features. Video-based Multi-view Graph Attention Network is to reveal the underlying complementary relations within appearance and motion by constructing appearance-specific and motion-specific graphs, and reveal the correlated relations between appearance and motion by constructing appearance-motion correlation and motion-appearance correlation graphs.
\subsubsection{Query Punishment Module}
Only query-related information is needed to infer the correct answer. Moreover, when perform multi-step reasoning, people have a propensity to pay attention to different textual parts of the question, especially the long and compositional questions. Therefore, we propose a Query Punishment Module to mimic human's step-to-step reasoning behavior, and filter out irrelevant visual features in each reasoning step. Specifically, based on previous word-level contextual word embedding $Q = \left(Q^1, Q^2, Q^3, \dots, Q^{L_q}\right)$ and initialized word embedding $W = \left(w^1, w^2, w^3, \dots, w^{L_q}\right)$, we employ a self-attention mechanism to obtain the current step's question feature $q^{\left(t\right)}_w$.

\allowdisplaybreaks[4]
\begin{align}
\left\{
\allowdisplaybreaks[4]
    \begin{aligned}
        w^i_q & = L2Norm\left(W_{1}Q^i\right) \\
        \alpha^i_q & = softmax(W_{2}w^i_q) \\
        q^{\left(t\right)}_w & = \sum_{i=1}^{L_q} \alpha^i_q w^i
    \end{aligned}
   \right.,
\end{align}
where $W_1 \in \mathbb{R}^{d \times d}$ and $W_2 \in \mathbb{R}^{d \times 1}$ are learnable parameters. 
After obtaining the question representation $q^{t}_w$ at the $t$-th iteration, we use it to filter out query-irrelevant visual features. Specifically, we first transform our question into the visual space with a fully-connected network, then dot-product operation is used to model the relevance of the current question and visual clip features. With gate mechanism, we obtain the query-guided mask value for each clip feature. Formally, the punishment mechanism is given by:
\begin{equation}
\left\{
    \begin{aligned}
        \beta_{a,i}^{\left(t\right)} & = \Tilde{V^i_a} W^{\left(t\right)}_a q^{\left(t\right)}_w \\
        \beta_{a}^{\left(t\right)} & = \sigma \left(\left[\beta_{a,1}^{\left(t\right)}, \dots, \beta_{a,N}^{\left(t\right)}\right]\right)\\
        \Tilde{E^{(t)}_a} & = \left[\beta_{a,1}^{\left(t\right)}\Tilde{V^1_a}, \dots, \beta_{a,N}^{\left(t\right)}\Tilde{V^N_a}\right]\\
    \end{aligned}
    \right.,
\end{equation}
\begin{equation}
\left\{
    \begin{aligned}
        \beta_{m,i}^{\left(t\right)} & = \Tilde{V^i_m} W^{\left(t\right)}_m q^{\left(t\right)}_w \\
        \beta_{m}^{\left(t\right)} & = \sigma
        \left(\left[\beta_{m,1}^{\left(t\right)}, \dots, \beta_{m,N}^{\left(t\right)}\right]\right)\\
        \Tilde{E^{(t)}_m} & = \left[\beta_{m,1}^{\left(t\right)}\Tilde{V^1_m}, \dots, \beta_{m,N}^{\left(t\right)}\Tilde{V^N_m}\right]\\
    \end{aligned}
    \right.,
\end{equation}
where $W^{\left(t\right)}_a \in \mathbb{R}^{300 \times d}$ and $W^{\left(t\right)}_m \in \mathbb{R}^{300 \times d}$ are learnable parameters, and $\sigma$ is the sigmoid function. After multiplying the masks and corresponding clip-based appearance and motion features, we filter out current irrelevant visual information at each step. 
\begin{figure}[t]
    \centering
    \includegraphics[scale=0.15]{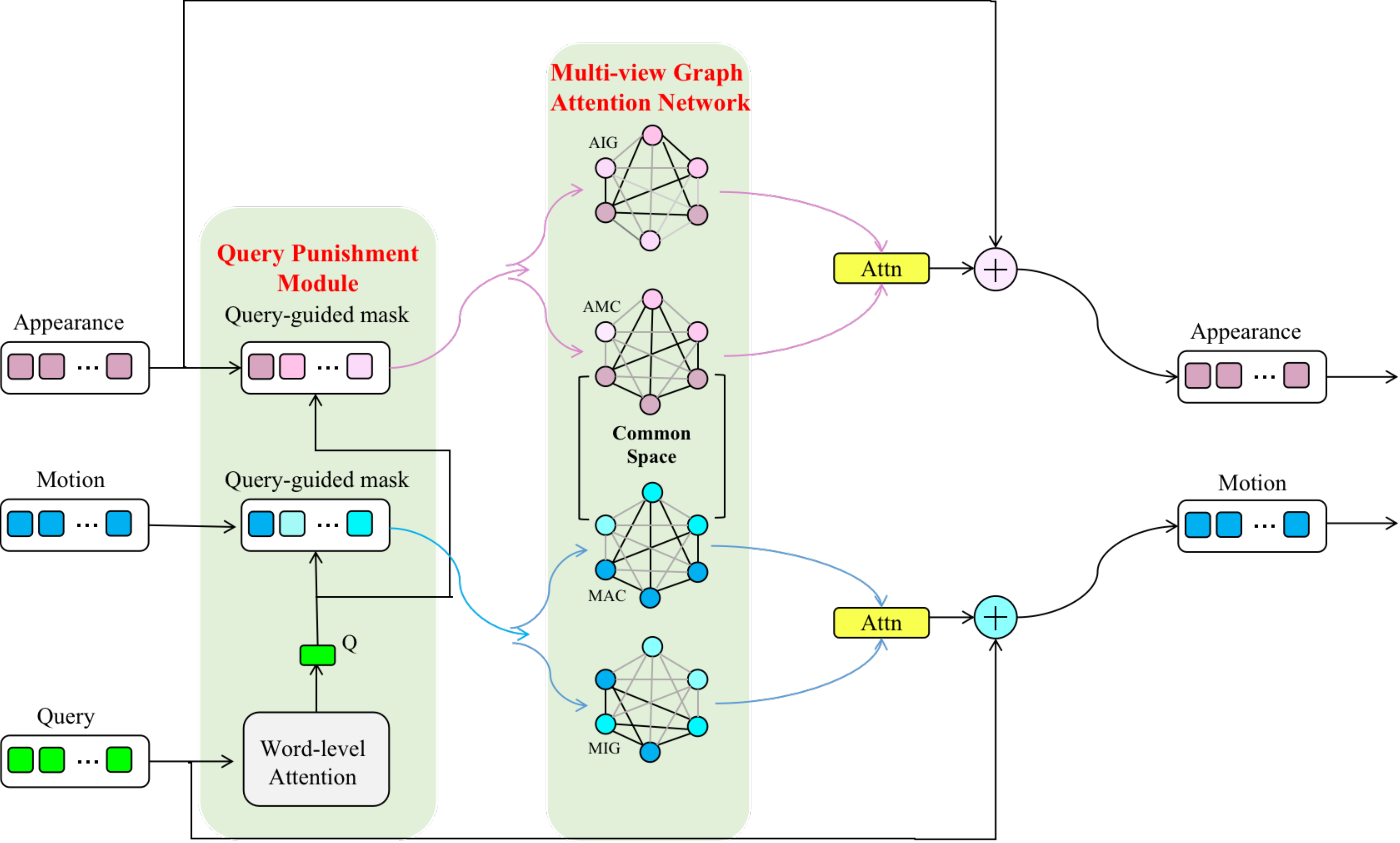}
    \caption{\textbf{DualVGR Unit.} DualVGR efficiently represents videos as amalgam of complementing information, including appearance, motion and relation features. The cell separately processes the appearance features and the motion features. First, query punishment module is implemented to filter out the irrelevant clips' features. Then, a multi-view graph network is used to provide a context-aware feature representation. With skip-connection mechanism, we get a combination of raw visual feature and visual relation feature. The unit's output is sent to another unit as the input feature representation.}
    \label{fig:DualVGR}
\end{figure}
\subsubsection{Video-based Multi-view Graph Attention Network}
As mentioned before, both appearance and motion channels are crucial for video understanding. In order to fully reveal the complemented information from these two channels, we need not only to extract the appearance and motion features from video clip itself, but also to consider the relations among video clips within each channel as well as the relations between two channels for each video clip. To this end, our work tries to aggregate the within-channel and between-channel relations into appearance and motion features. Inspired by GAT \cite{velivckovic2017graph}, which updates node representation over its neighbors with self-attention mechanisms, and has the ability to deal with both transductive learning and inductive learning problems, especially arbitrarily structured graph learning problems, we construct appearance graph with node as appearance feature of each clip, and motion graph with node as motion feature of each clip, then follow the proposed self-attention strategy in GAT to encode the neighborhood-relation into node representation for these two graphs. In this way, the within-channel relations are aggregated into appearance and motion features respectively. For between-channel relations, we follow the idea of AM-GCN \cite{wang2020gcn}, which is a new type of GCNs for graph classification task that can optimally integrate node features and topological structures by adaptively fusing specific and common embeddings from node features, topological structures and their combinations rather than simple GNN whose capability in extracting deep correlation information between topological structures and node features is distant from optimal, to seek one specific embedding and one common embedding for each channel by performing graph convolution operation. The specific embedding of one channel remains the specific characters of this channel, which should be disparity from that of the other channel. Two common embeddings model the correlated information from both channels, thus they should be consistency with each other.

For appearance channel, we construct two undirected complete graph networks, including appearance independent graph (AIG) and appearance-motion correlation graph (AMC), with each query punished clip-based appearance feature as a node. AIG aims to learn specific spatial-temporal contextual embeddings $Z_A^{\left(t\right)}$ within appearance channel. AMC is designed to extract the correlated relation features $Z_{CA}^{\left(t\right)}$ shared by appearance and motion channels. For motion channel, we also utilize two graph networks to learn the contextual embeddings, including motion independent graph (MIG) and motion-appearance correlated graph (MAC). Their goals are the same with graphs in appearance channel. According to these two motion graphs, we can get the specific embedding $Z_M^{\left(t\right)}$ and the correlated embedding $Z_{CM}^{\left(t\right)}$. Each graph is implemented with a multi-head Graph Attention Network (GAT) to model the relations between clip-features. Specifically, the attention score of each head $\alpha$ between two nodes is given by:

\begin{equation}
\left\{
    \begin{aligned}
        h^{t}_i & = U_t \Tilde{E^{\left(t\right)}_i} + b_t\\
        \alpha_{i,j}^{\left(t\right)} & = \frac{exp\left(LeakyReLU\left(W_t\left[h^{t}_i||h^{t}_j\right]\right)\right)}{\sum_{k \in N_i} exp\left(LeakyReLU\left(W_t\left[h^{t}_i||h^{t}_k\right]\right)\right)}
    \end{aligned}
    \right.,
\end{equation}

where $U_t \in \mathbb{R}^{d \times d_1}$ and $W_t \in 
\mathbb{R}^{2d_1 \times 1}$ are learnable parameters, $||$ is the concatenation operation, and $N_i$ is the first-order neighborhood of node $i$ in the graph.

Once the multi-head attention scores for each node are obtained, we update the representation of each node by:
\begin{equation}
    \begin{aligned}
        \Tilde{h^{t}_i} & = \overset{K}{\underset{k=1}{||}} \sigma \left(\sum_{j \in N_i} \alpha_{i,j}^{k,t}h_j\right),\\
    \end{aligned}
\end{equation}
where $||$ represents concatenation, $\alpha_{i,j}^k$ are the normalized attention coefficients from the k-th attention mechanism $\alpha^k$. The final shape of each node feature is $\mathbb{R}^{Kd_1 \times 1}$. 

Finally, we propose two losses to enhance the multi-view representation ability of our graphs, that is, consistency constraint and disparity constraint. Consistency constraint is to enhance the commonality between the correlated contextual information of appearance and motion spaces. Disparity constraint to enhance the independence of specific embeddings and correlated embeddings. The formulations of these two constraints are listed as follows.

\textbf{Consistency Constraint}:
For two output embeddings $Z_{CA}$ and $Z_{CM}$, we design a consistency constraint for our multi-view task, which could enhance their commonality. We first normalize the final embedding matrix $Z_{CA}^{\left(t\right)}$ and $Z_{CM}^{\left(t\right)}$ into $Z_{CAnor}^{\left(t\right)}$ and $Z_{CMnor}^{\left(t\right)}$ respectively. For simplicity, here we use $Z_{CAnor}$ to represent $Z_{CA}^{\left(t\right)}$, and $Z_{CMnor}$ to represent $Z_{CM}^{\left(t\right)}$.
Then, the similarity of $N$ embeddings can be calculated by:
\begin{equation}
    \begin{aligned}
    S_A^{\left(t\right)} & = Z_{CAnor} \cdot Z_{CAnor}^{\mathrm{T}} \\
    S_M^{\left(t\right)} & = Z_{CMnor} \cdot Z_{CMnor}^{\mathrm{T}}
    \end{aligned}
\end{equation}
The consistency indicates that the two similarity matrices should be similar as much as possible. The consistency loss of a single unit can be represented as:
\begin{equation}
    \mathcal{L}_c^{\left(t\right)} = \parallel S_A^{\left(t\right)} - S_M^{\left(t\right)} \parallel_2.
\end{equation}
Then the consistency loss of the whole network is:
\begin{equation}
    \mathcal{L}_c = \frac{1}{T} \sum^{t=1}_{T}{\mathcal{L}_c^{\left(t\right)}}, \label{eq:consistency}
\end{equation}
where $T$ is the number of iteration steps.

\textbf{Disparity Constraint}:
Since specific embeddings and common embeddings, such as $Z_{A}$ and $Z_{CA}$, are learned from graphs with the same topology, to make sure that these two embeddings could capture different information, we employ the Hilbert-Schmidt Independence Criterion (HSIC) \cite{song2007supervised} to enhance the independence of these two embeddings. HSIC, a simple yet effective measure of independence, has been implemented to several machine learning tasks \cite{gretton2005measuring}. For appearance, we use $Z_{A}$ to represent $Z_{A}^{\left(t\right)}$ and $Z_{CA}$ to represent $Z_{CA}^{\left(t\right)}$ for simplicity. The HSIC constraint of $Z_{A}^{\left(t\right)}$ and $Z_{CA}^{\left(t\right)}$ in a single unit is defined as:
\begin{equation}
    HSIC^{\left(t\right)}\left(Z_{A},Z_{CA}\right) = \left(n-1\right)^{-2}tr(RK_ARK_{CA}),
\end{equation}
where $K_A$ and $K_{CA}$ are the Gram matrices with $k_{A,ij} = k_A\left(z^i_A,z^j_A\right)$ and $k_{CA,ij} = k_{CA}\left(z^i_{CA},z^j_{CA}\right)$. $R = I - \frac{1}{n}ee^\mathrm{T}$, where $I$ is an identity matrix and $e$ is an all-one column vector. Follow AM-GCN \cite{wang2020gcn}, we use the inner product kernel function for $K_A$ and $K_{CA}$.

Similarly, the disparity constraint of embeddings $Z_{M}^{\left(t\right)}$ and $Z_{CM}^{\left(t\right)}$ in motion space can be given by:
\begin{equation}
    HSIC^{\left(t\right)}\left(Z_{M},Z_{CM}\right) = \left(n-1\right)^{-2}tr(RK_MRK_{CM}).
\end{equation}

Finally, the disparity constraint for the whole network is given by:
\begin{gather}
\mathcal{L}_d^{t} = HSIC^{\left(t\right)}\left(Z_{A},Z_{CA}\right)+HSIC^{\left(t\right)}\left(Z_{M},Z_{CM}\right), \\
\mathcal{L}_d = \frac{1}{T}\sum^{t=1}_{T}\left(\mathcal{L}_d^{t}\right), \label{eq:disparity}
\end{gather}
where $T$ is the number of iteration steps.

\subsubsection{Attention} \label{sec:Attn}
Now we have two embeddings $Z_A$ and $Z_{CA}$ in appearance space, and two embeddings $Z_{M}$ and $Z_{CM}$ in motion space. In order to adaptively capture the contextual information for each space, we utilize attention mechanism to fuse two embeddings in appearance space, as well as those in motion space. Their corresponding attention importance scores $\left(\alpha_a, \alpha_{ca}\right)$ and $\left(\alpha_m, \alpha_{cm}\right)$ are given by:
\begin{align}
    \left(\alpha_a, \alpha_{ca}\right) & = Attn\left(Z_A, Z_{CA}\right) \label{eq:attn},\\
    \left(\alpha_m, \alpha_{cm}\right) & = Attn\left(Z_M, Z_{CM}\right) \label{eq:attn2}.
    \end{align}
The embedding of the $i$-th clip in appearance space in $Z_A$ is $z^{\left(i,t\right)}_A \in \mathbb{R}^{Kd_1 \times 1}$ and the embedding of the $i$-th clip in motion space in $Z_M$ is $z^{\left(i,t\right)}_M \in \mathbb{R}^{Kd_1 \times 1}$. First, we transform the embeddings through a nonlinear transformation $w^{\left(t\right)}_a \in \mathbb{R}^{Kd_1 \times Kd_1}$ and $w^{\left(t\right)}_m \in \mathbb{R}^{Kd_1 \times Kd_1}$ respectively, then use a weight vector $U^{\left(t\right)}_a \in \mathbb{R}^{Kd_1 \times 1}$ and $U^{\left(t\right)}_m \in \mathbb{R}^{Kd_1 \times 1}$ to get the attention value $v^i_A$ and $v^i_M$ for the $i$-th clip as follows:
\begin{equation}
\left\{
\begin{aligned}
    v^{\left(i,t\right)}_A = U^{\left(t\right)}_a tanh(w^{\left(t\right)}_a z^{\left(i,t\right)}_A + b_a),\\
    v^{\left(i,t\right)}_M = U^{\left(t\right)}_m tanh(w^{\left(t\right)}_m z^{\left(i,t\right)}_M + b_m).
\end{aligned}
\right.
\end{equation}
Similarly, we can get the attention score $v^{\left(i,t\right)}_{CA}$ for $Z_{CA}$ and the attention score $v^{\left(i,t\right)}_{CM}$ for $Z_{CM}$. Then, we normalize the attention values with softmax function to get the final score:
\begin{equation}
\left\{
\begin{aligned}
    \alpha^{\left(i,t\right)}_A = \frac{exp\left(v^{\left(i,t\right)}_{A}\right)}{exp\left(v^{\left(i,t\right)}_{A}\right) + exp\left(v^{\left(i,t\right)}_{CA}\right)}, \\
    \alpha^{\left(i,t\right)}_M = \frac{exp\left(v^{\left(i,t\right)}_{M}\right)}{exp\left(v^{\left(i,t\right)}_{M}\right) + exp\left(v^{\left(i,t\right)}_{CM}\right)}.
    \end{aligned}
    \right.
\end{equation}
Finally, we obtain the final embedding $z^{\left(i,t\right)}_a$ in appearance space by combining these two embeddings $z^{\left(i,t\right)}_A$ and $ z^{\left(i,t\right)}_{CA}$, and the final embedding $z^{\left(i,t\right)}_m$ in motion space by combining $z^{\left(i,t\right)}_M$ and $ z^{\left(i,t\right)}_{CM}$:
\begin{equation}
\left\{
\begin{aligned}
    z^{\left(i,t\right)}_a = \alpha^{\left(i,t\right)}_A \cdot z^{\left(i,t\right)}_A + \alpha^{\left(i,t\right)}_{CA} \cdot z^{\left(i,t\right)}_{CA}, \\
    z^{\left(i,t\right)}_m = \alpha^{\left(i,t\right)}_M \cdot z^{\left(i,t\right)}_M + \alpha^{\left(i,t\right)}_{CM} \cdot z^{\left(i,t\right)}_{CM}.
    \end{aligned} \label{eq:attn_final}
    \right.
\end{equation} 
Then, a residual connection \cite{he2016deep} is used to avoid the vanishing gradient problem. This operation can also be considered as an amalgam of complementing factors including appearance, motion and query-related visual relation information. Each clip-based feature is updated by:
\begin{equation}
    \left\{
\begin{aligned}
    \Tilde{V^i_a} = \Tilde{V^i_a} + z^{(i,t)}_a, \\
    \Tilde{V^i_m} = \Tilde{V^i_m} + z^{(i,t)}_m.
    \end{aligned}
    \right.
\end{equation}

\subsubsection{Multi-step Reasoning} 
Finally, the DualVGR unit is stacked as a chain to perform the final DualVGR network:
\begin{equation}
    \left\{\Tilde{V^i_a},\Tilde{V^i_m}\right\} = DualVGR(\Tilde{V^i_a};\Tilde{V^i_m};Q;W).
\end{equation}

Through multiple steps of iteration, our DualVGR's final visual representation contains complementary information about the question-related clips, including appearance, motion and the corresponding relations between them. 

\subsection{Video Representation Fusion}
At the final step, we obtain the appearance and motion features for each clip $\Tilde{V^i_a}$ and $\Tilde{V^i_m}$, we use Multimodal Factorized Bilinear pooling (MFB) \cite{yu2017multi} to fuse them to get the final visual representation of each clip $V^i_f$. An attention mechanism called Graph Readout Operation \cite{wu2020comprehensive} is used to fuse the clip features. The details are illustrated as follows:
\begin{align}
    V^i_f & = MFB(\Tilde{V^i_a}, \Tilde{V^i_m}), \\
    \Tilde{o} & = Attn\left(V^i_f\right), 
\end{align}
where $\Tilde{o} \in \mathbb{R}^{d \times 1}$ is the output visual feature. The attention mechanism is implemented as Eqn. \eqref{eq:attn} - Eqn. \eqref{eq:attn_final}.

\subsection{Answer Decoder}
Following \cite{le2020hierarchical}, we adopt the same answer decoders of these open-ended questions for the fair comparison:
\begin{align}
    y & = ELU\left(W_o\left[\Tilde{o},W_QE^Q+b\right]+b\right), \\
    y' & = ELU\left(W_yy+b\right), \\
    p & = softmax\left(W_{y'}y'+b\right), \label{eq:prob}
\end{align}
where $W_Q \in \mathbb{R}^{d \times d}$ and $W_o \in \mathbb{R}^{2d \times d}$.
\subsection{Total Loss}\label{sec:Loss}
We cast our open-ended VideoQA task as a classification task. Hence, we use cross-entropy loss $\mathcal{L}_t$ for this task. Then, the total loss is:
\begin{equation}
    \mathcal{L} = \mathcal{L}_t + \gamma \mathcal{L}_c + \beta \mathcal{L}_d,
\end{equation}
where $\gamma$ and $\beta$ are parameters of the consistency and disparity constraint terms. The consistency constraint is implemented as Eqn. \eqref{eq:consistency} and disparity constraint is implemented as Eqn. \eqref{eq:disparity}.


%
\section{Experiments} \label{sec:experiments}
\begin{table}[t]
    \centering
    \caption{Statistics of the MSVD-QA Dataset.}
    \resizebox{0.45\textwidth}{!}{
    \begin{tabular}{cccccccc}
    \hline
        \multirow{2}{*}{\qquad} & \multirow{2}{*}{Video} & \multirow{2}{*}{QA pair} & \multicolumn{5}{c}{Question Type}\\\cline{4-8}  
        &&&what&who&how&when&where \\\hline
        Train&1,200&30,933&19,485&10,469&736&161&72\\
        Val&250&6,415&3,995&2,168&185&51&16\\
        Test&520&13,157&8,149&4,552&370&58&28\\\hline
        All&1,970&50,505&31,629&17,199&1,291&270&116\\\hline
    \end{tabular}}
    \label{tab:MSVD-QA}
\end{table}

\begin{table}[t]
    \centering
    \caption{Statistics of the MSRVTT-QA Dataset.}
    \resizebox{0.45\textwidth}{!}{
    \begin{tabular}{cccccccc}
    \hline
        \multirow{2}{*}{\qquad} & \multirow{2}{*}{Video} & \multirow{2}{*}{QA pair} & \multicolumn{5}{c}{Question Type}\\\cline{4-8}  
        &&&what&who&how&when&where \\\hline
        Train&6,513&158,581&108,792&43,592&4,067&1,626&504\\
        Val&497&12,278&8,337&3,439&344&106&52\\
        Test&2,990&72,821&49,869&20,385&1,640&677&250\\\hline
        All&10,000&243,680&166,998&67,416&6,051&2,409&806\\\hline
    \end{tabular}}
    \label{tab:MSRVTT-QA}
\end{table}
\subsection{Datasets}
1) \textit{MSVD-QA} \cite{xu2017video}: There are $1970$ trimmed videos collected from the Microsoft Research Video Description (MSVD) Corpus \cite{chen2011collecting}. It contains 50,500 QA pairs automatically generated by the NLP algorithm in total, which contains five general types of questions, including what, how, when, where and who. The average video length is approximately 10 seconds, and the average question length is approximately 6 words. Therefore, it is a small-scale dataset with short questions. We conduct experiments on this dataset to test our model’s generalization ability in short videos in real worlds. The details of MSVD-QA dataset are illustrated in Table \ref{tab:MSVD-QA}.

2) \textit{MSRVTT-QA} \cite{xu2016msr}: Compared with MSVD-QA, MSR-VTT contains longer videos and more complex scenes. We conduct experiments on this dataset to test our model’s performance for longer videos of real datasets. It contains 10,000 trimmed videos from MSR-VTT dataset \cite{xu2016msr} and 243,000 QA pairs generated by the NLP algorithm. The average video length is approximately 15 seconds, and the average question length is approximately 7 words. The details of MSRVTT-QA are illustrated in Table \ref{tab:MSRVTT-QA}.

3) \textit{SVQA} \cite{song2018explore}:
It is a large-scale synthetic dataset which contains 12,000 synthetic videos and around 120K QA pairs. Specifically, videos are generated from Unity3D, and each video length is the same as 10 seconds. Meanwhile, QA pairs are generated from question templates automatically, and the questions are generated exclusively long with an average length of 20 words. Furthermore, each question can be decomposed into human readable logical chain or tree layout easily. The goal of this dataset is to test the reasoning ability of VideoQA systems. Table \ref{tab:SVQA} illustrates the statistics of SVQA dataset.

\begin{table}[t]
    \centering
    \caption{Statistics of the SVQA Dataset.}
    \resizebox{0.45\textwidth}{!}{
    \begin{tabular}{l|c|ccc}
    \hline
         Question Category&Sub Category&Train&Val&Test  \\\hline\hline
         Count&$/$&19,320&2,760&5,520\\\hline
         Exist&$/$&6,720&960&1,920\\\hline
         \multirow{5}{5em}{Query}&Color&7,560&1,056&2,160\\
         &Size&7,560&1,056&2,160\\
         &Action Type&6,720&936&1,920\\
         &Direction&7,560&1,056&2,160\\
         &Shape&7,560&1,056&2,160\\\hline
         \multirow{3}{3em}{Integer Comparison}&More&2,520&600&720\\
         &Equal&2,520&600&720\\
         &Less&2,520&600&720\\\hline
         \multirow{5}{5em}{Attribute Comparison}&Color&2,520&216&720\\
         &Size&2,520&216&720\\
         &Action Type&2,520&216&720\\
         &Direction&2,520&216&720\\
         &Shape&2,520&216&720\\\hline
         \multicolumn{2}{l}{Total QA pairs}&83,160&11,880&23,760\\\hline
         \multicolumn{2}{l}{Total Videos}&8,400&1,200&2,400\\\hline
    \end{tabular}}
    \label{tab:SVQA}
\end{table}

\subsection{Implementation Details}
For each video in MSVD-QA, we segment the video into $8$ clips. For MSRVTT-QA, each video is divided into $16$ clips. Besides, in SVQA dataset, each video is splitted into $20$ clips. The numbers of divided clips are determined by grid search method from $\{4,8,12,16,20,24\}$. For all the datasets, each clip contains $16$ frames by default. The video appearance and question encoders are one-layer BiLSTMs. The dimension $d=768$, $d1=192$, and the number of the multi-heads $k$ is $4$. The iteration step $t$ for MSVD-QA is set to $1$, the iteration step $t$ for MSRVTT-QA is $6$, and the iteration step $t$ for SVQA is $4$. For loss function, we use the grid search method to select the best coefficients. Specifically, the value set of $\gamma$ is $\{0,1e-4,1e-3,1e-2,1e-1,1,10,100,1000\}$, while the value set of the remaining coefficient $\beta$ is $\{0,1e-8,1e-6,1e-4,1e-2,1,10,100\}$. After searching, we obtain the best coefficients for all the datasets: $\gamma$ = $100$, $\beta$ = $1e-6$ in MSVD-QA, $\gamma$ = $100$, $\beta$ = $1e-6$ in MSRVTT-QA and $\gamma$ = $1$, $\beta$ = $1e-6$ in SVQA. Our framework is implemented in PyTorch, and the network is trained by Adam optimizer with a fixed learning rate $1e-4$. The batch size is set to $256$. All experiments are terminated after $25$ epochs and the results are reported at the epoch which has the best validation accuracy.

\subsection{Comparison with the State-of-the-art}
We compare our proposed model with state-of-the-art methods (SOTA) on aforementioned datasets. For MSVD-QA and MSRVTT-QA, we compare with most resent SOTA, including HME \cite{fan2019heterogeneous}, HGA \cite{jiang2020reasoning}, HCRN \cite{le2020hierarchical} and TSN \cite{yang2019question}.
\begin{itemize}
    \item HME is a model equipped with memory network. It uses a redesigned question memory to improve the question representation in each step. The visual representations, including appearance and motion features, are mapped into the heterogeneous memory. Then, multi-step reasoning is performed with self-updated attention in this memory network.
    \item HGA is a model with graph network. It represents all video shots and question words as the graph to perform cross-modal reasoning.
    \item HCRN is a model with stacked clip-based relation networks. The CRN takes input as an array of tensorial objects and a conditioning feature, and output as an array of relation information within them. By hierarchically stacking these blocks, HCRN performs multi-step relational reasoning.
    \item TSN is a model containing several modules to perform multi-step reasoning. For example, since appearance and motion play different roles in multi-step reasoning, a switch module has been proposed to adaptively choose appearance or motion channel as the primary channel, guiding the reasoning process. 
\end{itemize}

The results are summarized in Table \ref{tab:SoTAmsvd-msrvtt} for MSVD-QA and MSRVTT-QA. It is clear that our framework consistently outperforms or is competitive with SOTA models on all tasks for short questions (average question length $\leq 7$ words). For MSVD-QA dataset, our DualVGR achieves $39.03\%$ overall accuracy, which is $2.33\%$ improvement over previous SOTA methods. It also achieves quite high performance in each question types. For instance, as shown in Table \ref{tab:SoTAmsvd-msrvtt}, DualVGR achieves approximately 3\% improvement for the question of ``what'' and ``who''. The improvement of DualVGR compared with HME and TSN confirms the effectiveness of relational reasoning in VideoQA tasks. Furthermore, the improvement of DualVGR compared with HGA shows that the critical role of query punishment module in our framework.
\begin{table*}[t]
    \centering
    \caption{Performance comparison to the state-of-the-art methods on MSVD-QA and MSRVTT-QA.}
    \begin{tabular}{ccccccccccccc}
    \hline
        \multirow{2}{*}{\qquad} & \multicolumn{6}{c}{MSVD-QA} & \multicolumn{6}{c}{MSRVTT-QA} \\\cmidrule(lr){2-7} \cmidrule(lr){8-13}
        &what&who&how&when&where&ALL&what&who&how&when&where&ALL\\\hline
        HME \cite{fan2019heterogeneous}&22.4&50.1&73.0&70.7&42.9&33.7&26.5&43.6&82.4&76.0&28.6&33.0 \\
        HGA \cite{jiang2020reasoning}&23.5&50.4&83.0&72.4&46.4&34.7&29.2&45.7&83.5&75.2&34.0&35.5 \\
        HCRN \cite{le2020hierarchical}&/&/&/&/&/&36.1&/&/&/&/&/&\textbf{35.6}\\
        TSN \cite{yang2019question}&25.0&51.3&\textbf{83.8}&\textbf{78.4}&\textbf{59.1}&36.7&27.9&\textbf{46.1}&\textbf{84.1}&\textbf{77.8}&\textbf{37.6}&35.4\\\hline
        Ours&\textbf{28.65}&\textbf{53.82}&80.00&70.69&46.43&\textbf{39.03}&\textbf{29.40}&45.56&79.76&76.66&36.40&35.52\\\hline
    \end{tabular}
    \label{tab:SoTAmsvd-msrvtt}
\end{table*}
\begin{table*}[t]
    \centering
    \caption{Performance comparison to the state-of-the-art methods on SVQA.}
    \resizebox{\textwidth}{!}{
    \begin{tabular}{ccccccccccccccccc}
    \hline
        \multirow{2}{*}{\qquad} & \multirow{2}{*}{Exist} & \multirow{2}{*}{Count} & \multicolumn{3}{c}{Integer Comparison} & \multicolumn{5}{c}{Attribute Comparison} & \multicolumn{5}{c}{Query} & \multirow{2}{*}{All} \\\cmidrule(lr){4-6} \cmidrule(lr){7-11} \cmidrule(lr){12-16}
        &&&More&Equal&Less&Color&Size&Type&Dir&Shape&Color&Size&Type&Dir&Shape&\\\hline
        Unified-Attn \cite{xue2017unifying}&\textbf{54.17}&35.23&67.05&52.98&50.00&51.85&54.26&50.14&50.28&49.43&22.01&53.73&56.51&34.70&38.20&41.69\\
        SA+TA-GRU \cite{song2018explore}&52.03&38.20&74.28&57.67&61.60&55.96&\textbf{55.90}&\textbf{53.40}&\textbf{57.50}&\textbf{52.98}&23.39&\textbf{63.30}&62.90&\textbf{43.20}&41.69&44.90\\
        STRN \cite{singh2019spatio}&54.01&44.67&72.22&\textbf{57.78}&\textbf{62.92}&56.39&55.28&50.69&50.14&50.00&24.31&59.68&59.32&28.24&44.49&47.58\\\hline
        Ours&54.15&\textbf{49.55}&\textbf{80.89}&57.63&59.71&\textbf{56.87}&51.90&49.79&49.78&51.53&\textbf{27.96}&60.65&\textbf{64.19}&34.46&\textbf{47.48}&\textbf{50.38}\\\hline
    \end{tabular}}
    \label{tab:SoTASVQA}
\end{table*}
For MSRVTT-QA, our model achieves $35.52\%$ accuracy, which is only $0.08\%$ lower than the current state-of-the-art performance. Compared with SOTA method HCRN which conducts the question-aware frame-level feature and the question-aware clip-level feature in a hierarchical structure, our method only extracts question-aware clip-level feature, which reduces the computational cost, with only a minimum drop of performance even in dataset with complicated scenes. To sum up, MSVD-QA and MSRVTT-QA results imply that our model can handle real-world videos well.

We further compare our methods with SOTA methods on synthetic dataset, SVQA. This dataset contains many compositional questions, which require agents to be able to perform multi-step relational reasoning to infer the answer. For SVQA, three video question answering models as well as ours are used for comparisons:

\begin{itemize}
    \item Unified-Attn \cite{xue2017unifying} is a model with two attention mechanisms, including sequential video attention and temporal question attention mechanisms. 
    \item SA+TA-GRU \cite{song2018explore} is a model with a refined GRU whose hidden state transfer process is associated with temporal attention to strengthen long-term temporal dependency.
    \item STRN \cite{singh2019spatio} is a model with spatio-temporal relational network, which aims to model temporal changes in both the interactions among different objects and the motion-dynamics of individual objects.
\end{itemize}

The results are summarized in Table \ref{tab:SoTASVQA}. From the results, we can observe that our proposed model DualVGR outperforms the state-of-the-art methods. Specifically, our framework achieves the best overall accuracy $50.38\%$, leading to approximately $2.8\%$ improvement over the best compared method STRN. Our method achieves promising results on all question types in SVQA dataset, especially $4.88\%$ improvement on ``Count'' questions. On the contrary, previous methods perform poorly on ``Count'' questions, which indicates the powerful generalization ability of our method for multi-step reasoning. The aforementioned results indicate the effectiveness of the DualVGR for long and compositional questions. Furthermore, the method STRN is implemented with relation network, which means that DualVGR unit with graph network may be more suitable for relational reasoning task in VideoQA. 
\subsection{Ablation Study}
To prove the effectiveness of the essential components of DualVGR unit and to provide more detailed parameter analysis, we conduct extensive ablation studies on MSVD-QA test set and SVQA test set with a wide range of configurations. The results of the component analysis are reported in Table \ref{tab:ablationDualVGR}.

\subsubsection{The Effectiveness of Essential Components}
The whole architecture of our DualVGR unit mainly incorporates two essential components: Query Punishment Module and Video-based Multi-view Graph Attention Network. We consider the following ablation models to verify the importance of each component in VideoQA:
\begin{itemize}
    \item \textbf{AG}: this model extracts appearance feature from video clips, and constructs them as the appearance graph. Then, multi-head GAT is implemented to get the contextual representation. At last, graph attention readout mechanism is used to get the final visual vector, and fuse it with the question vector to infer the answer.
    \begin{figure*}[t]
    \centering
    \subfigure[]{
    \centering
    \includegraphics[scale=0.36]{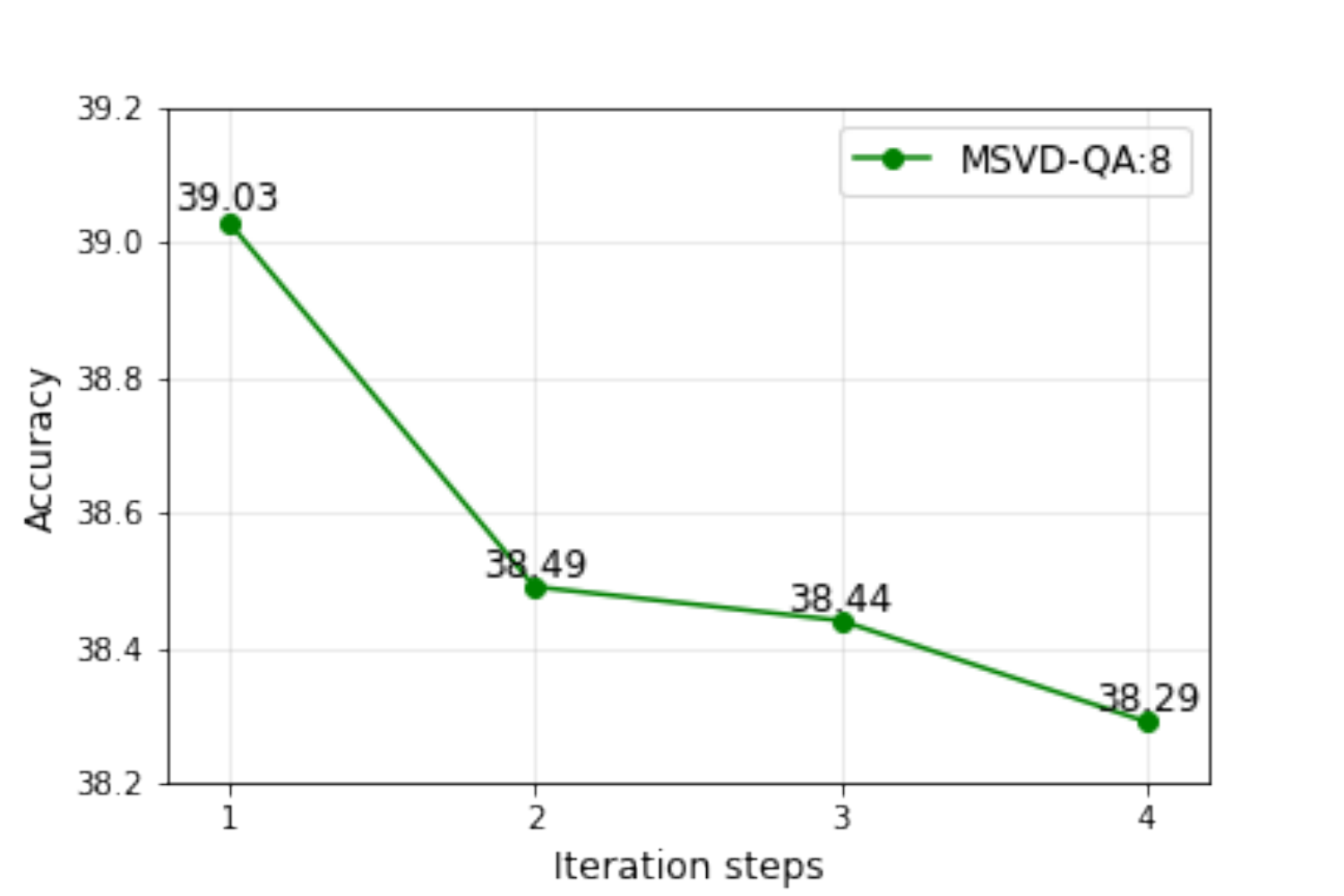}
    } 
    \subfigure[]{
    \centering
    \includegraphics[scale=0.36]{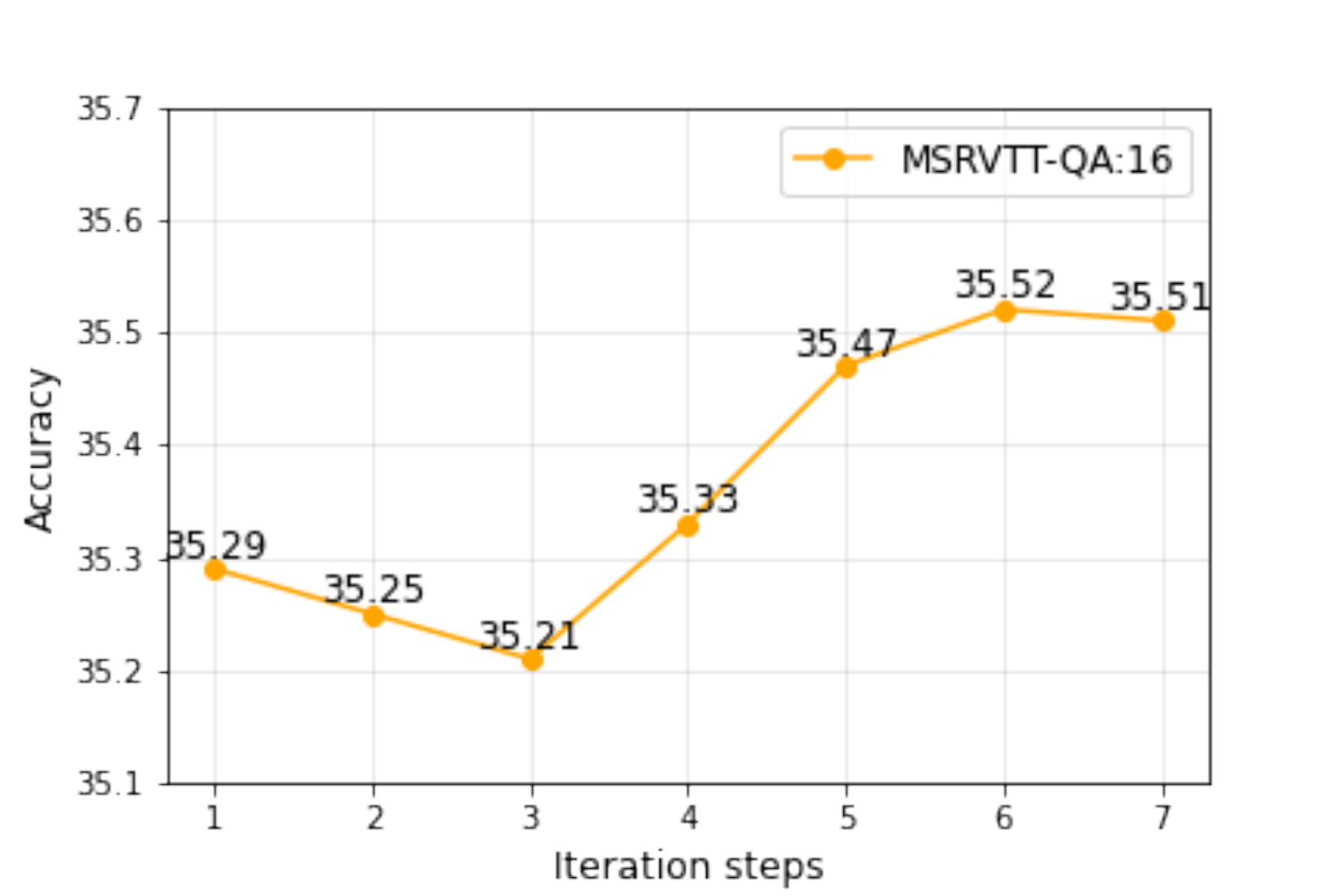}
    } 
    \subfigure[]{
    \centering
    \includegraphics[scale=0.36]{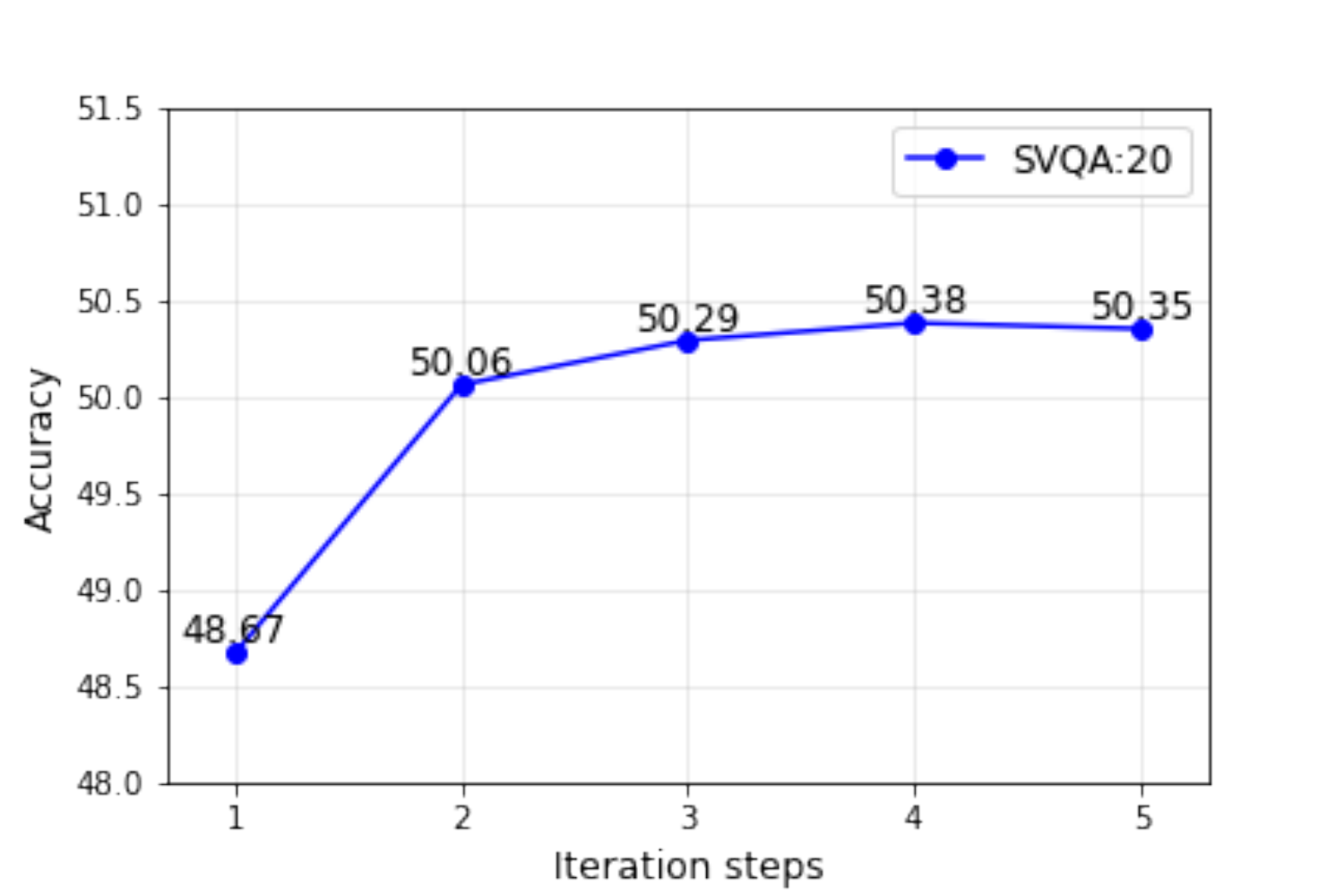}
    }
    
    \caption{\textbf{Number of iterations}. Impact of the number of steps in the iterative process on MSVD, MSRVTT-QA and SVQA}
    \label{fig:iteration}
\end{figure*}
    \item \textbf{MG}: compared with AG, this model extracts motion feature from video clips, and constructs them as the motion graph.
    \item \textbf{FG}: compared with AG, this model extracts both appearance and motion features from video clips, and fuses them into a new clip-based visual vector with MLP. Then, the visual graph is constructed with them. 
    \item \textbf{Bigraph}: compared with AG, this model constructs both appearance graph and motion graph. Then, we apply two multi-head GAT to capture contextual representations of both visual spaces in each unit. Next, appearance and motion features of each clip are fused with MFB.
    \item \textbf{MVgraph}: compared with Bigraph, this model uses Video-based Multi-view Graph Attention Network to learn the contextual representations of appearance and motion features.
    \item \textbf{PFG}: compared with FG, this model implements Query Punishment Module to keep the relevant visual features. Then, multi-head GAT is utilized to represent the contextual information.
    \item \textbf{DualVGR}: this is our proposed model. Compared with MVgraph, this model adds Query Punishment Module in our unit.
    \item \textbf{sharedVGR}: compared with DualVGR, this model use one shared multi-head GAT to learn the contextual representations of both visual spaces, including appearance and motion spaces.
\end{itemize}

The results are shown in Table \ref{tab:ablationDualVGR}, and the key observations and conclusions are as follows: (1) \textbf{Two-stream Visual Features}: \textbf{AG} and \textbf{MG} just consider one type of visual information, such as appearance and motion features. \textbf{FG} combines the appearance and motion features into a new visual space, and learn the contextual representations of this space. After considering the appearance and motion features, it increases by $1.91\%$ and $0.26\%$ on SVQA dataset, and $0.45\%$ and $3.83\%$ on MSVD-QA dataset. This illustrates the effectiveness of utilizing two-stream visual features. Furthermore, as we can see, AG outperforms MG in MSVD-QA and MG outperforms AG in SVQA. This can be attributed to the differences between MSVD-QA and SVQA. For instance, questions in SVQA always require agents to understand the spatial-temporal relationships between all clips, hence motion features might be more important than appearance features. Questions in MSVD-QA are usually very simple, which only require agents to analyze certain clips to answer the questions. Therefore, motion features would be less important in MSVD-QA. (2) \textbf{Multi-view Graph Attention Network}: \textbf{Bigraph} utilizes two multi-head GAT to learn the contextual representations of visual features. \textbf{MVgraph} implements our Multi-view Graph Attention Network to learn the contextual representations. The results imply that our Multi-view Graph Attention Network is more suitable for this task, especially multi-view relation learning. However, we observe that MVgraph underperforms the FG in both datasets. The detailed analysis will be illustrated later. (3) \textbf{Query Punishment Module}: \textbf{PFG} further considers Query Punishment Module in our FG. \textbf{DualVGR} is our whole unit with Query Punishment Module and Video-based Multi-view Graph Attention Network. After the consideration of involving Query Punishment Module to our unit, it increases $0.51\%$ and $1.53\%$ on SVQA dataset, and $0.29\%$ and $1.9\%$ on MSVD-QA dataset, which proves the advantage of our Query Punishment Module to filter our irrelevant information. Moreover, DualVGR outperforms PFG, which is distinct from what was discussed above. This could be putted down to that MVgraph and FG just represent all video clips as useful node to propogate information with GAT which could lead to much noise. Consequently, with more graphs in MVgraph, it contains more noise than FG, which leads to worse performance. With Query Punishment Module, our DualVGR successfully achieves better performance than PFG. \textbf{shareDVGR} only uses one shared Multi-head GAT to extract the common contextual information in both visual spaces. We can observe from the results that with two common graphs in the unit, we obtain better results than one shared graph. (4) \textbf{Self-Attention for Questions}: \textbf{simpleDualVGR} is the DualVGR framework without self-attention mechanism in each reasoning step. We can observe that with self-attention mechanism, our DualVGR obtains $1.36\%$ performance gain in SVQA dataset. It verify the effectiveness of self-attention mechanism when performing multi-step reasoning for long and compositional questions.
\begin{table}[t]
    \centering
    \caption{Ablation studies on MSVD-QA and SVQA about essential components of DualVGR.}
     \resizebox{0.45\textwidth}{!}{
    \begin{tabular}{c|c|c|cccccc}
    \hline
        \multirow{2}{*}{Model} & \multirow{2}{*}{Step \textit{T}} & \multirow{2}{*}{SVQA} & \multicolumn{6}{c}{MSVD-QA}\\\cline{4-9}  &&&what&who&how&when&where&ALL\\\hline\hline
        \textbf{Two-stream}&&&&&&\\
        AG&\textit{T} = 1&46.54&26.97&52.17&79.46&\textbf{74.14}&\textbf{50.00}&37.42\\
        MG&\textit{T} = 1&48.19&23.11&49.56&77.30&70.69&46.43&34.04\\
        FG&\textit{T} = 1&48.45&27.77&51.85&\textbf{82.97}&67.24&\textbf{50.00}&37.87\\\hline
        \textbf{Multi-view}&&&&&&\\
        Bigraph&\textit{T} = 1&47.53&26.09&53.10&77.57&70.69&35.71&37.10\\
        MVgraph&\textit{T} = 1&47.59&26.75&52.07&75.68&72.41&46.43&37.13\\\hline
        \textbf{Punishment}&&&&&&\\
        PFG&\textit{T} = 1&48.96&27.50&53.87&74.59&67.24&46.43&38.16\\
        shareDVGR&\textit{T} = 1&48.96&27.94&52.35&77.84&72.41&\textbf{50.00}&38.03\\
        DualVGR&\textit{T} = 1&\textbf{49.12}&\textbf{28.65}&\textbf{53.82}&80.00&70.69&46.43&\textbf{39.03}\\\hline
        \textbf{self-attention}&&&&&&\\
        simpleDualVGR&\textit{T} = 4&49.02&/&/&/&/&/&/\\
        DualVGR&\textit{T} = 4&50.38&/&/&/&/&/&/\\\hline

    \end{tabular}}
    \label{tab:ablationDualVGR}
\end{table}
\begin{figure*}[ht]
    \centering
    \includegraphics[scale=0.3]{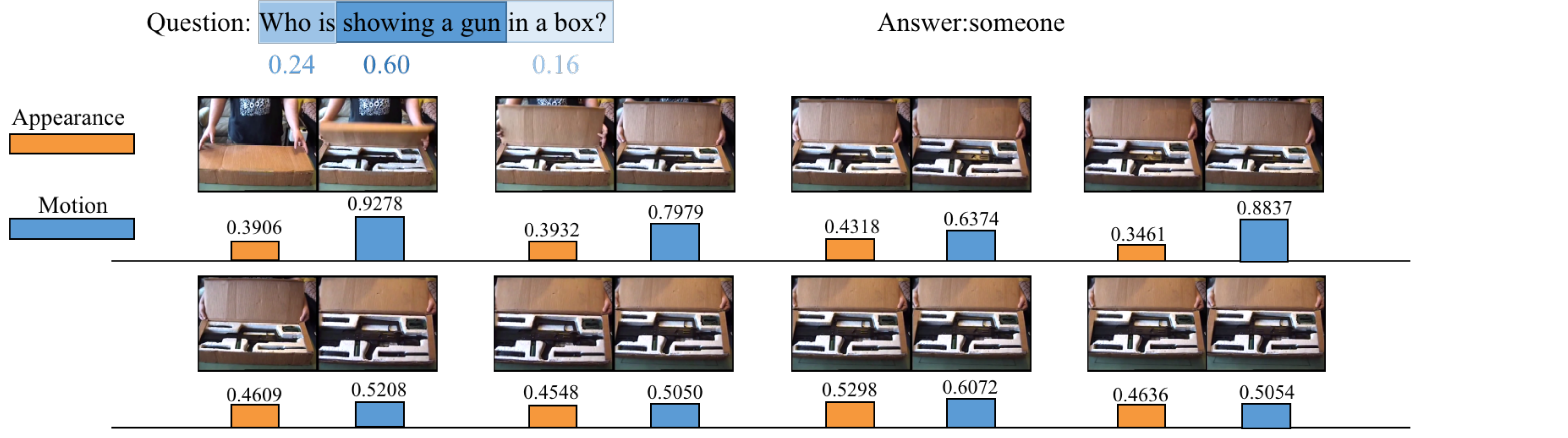}
    \end{figure*}
\begin{figure*}[ht]
    \centering
    \includegraphics[scale=0.3]{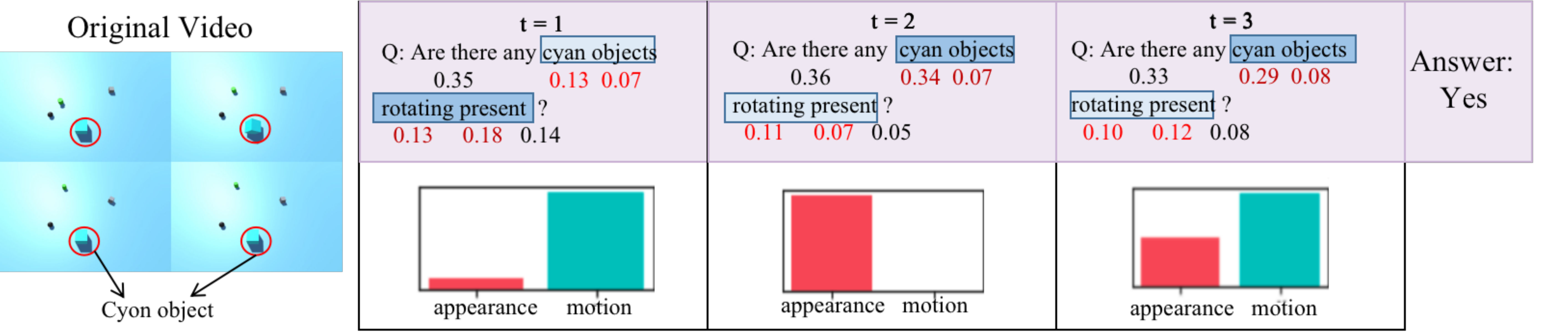}
    \caption{Visualization of the attention scores in each step's query representation and the generated query-guided visual mask of each visual feature in each step on MSVD-QA (the top row) and SVQA (the bottom row).}
    \label{fig:VisualEx}
\end{figure*}

\subsubsection{The detailed analysis of iteration steps $T$}
We perform the detailed analysis of the iterative process in Fig. \ref{fig:iteration}. We train our DualVGR network on all datasets with the configuration as our best model. First, questions in MSVD-QA dataset are usually very short, which implies that agents do not need to perform very complex relational reasoning to fully understand the questions. Consequently, the overall accuracy on MSVD-QA decreases when performing more steps of relational reasoning. Next, the test accuracy on MSRVTT-QA dataset increases from $35.29\%$ to $35.52\%$ when the number of reasoning iteration increases from $1$ to $6$. Since MSRVTT-QA dataset has more complex questions and longer videos than MSVD-QA, it could benefit from a higher number of reasoning iterations over our DualVGR unit. Then, when the iteration step is set to $7$, the model's performance becomes very stable. At last, SVQA is a dataset designed for testing the relational reasoning ability of models. Therefore, the questions are compositional and complex which require agents to perform multi-step relational reasoning. The results reveal that the accuracy of SVQA dataset also benefits from a higher number of iteration steps. Network with four steps provides a gain of $+1.71\%$ in overall test accuracy on SVQA over the network with a single step. Then, when the step increases by $5$, the model's performance becomes very stable. 
\begin{figure}[t]
    \centering
    \includegraphics[scale=0.4]{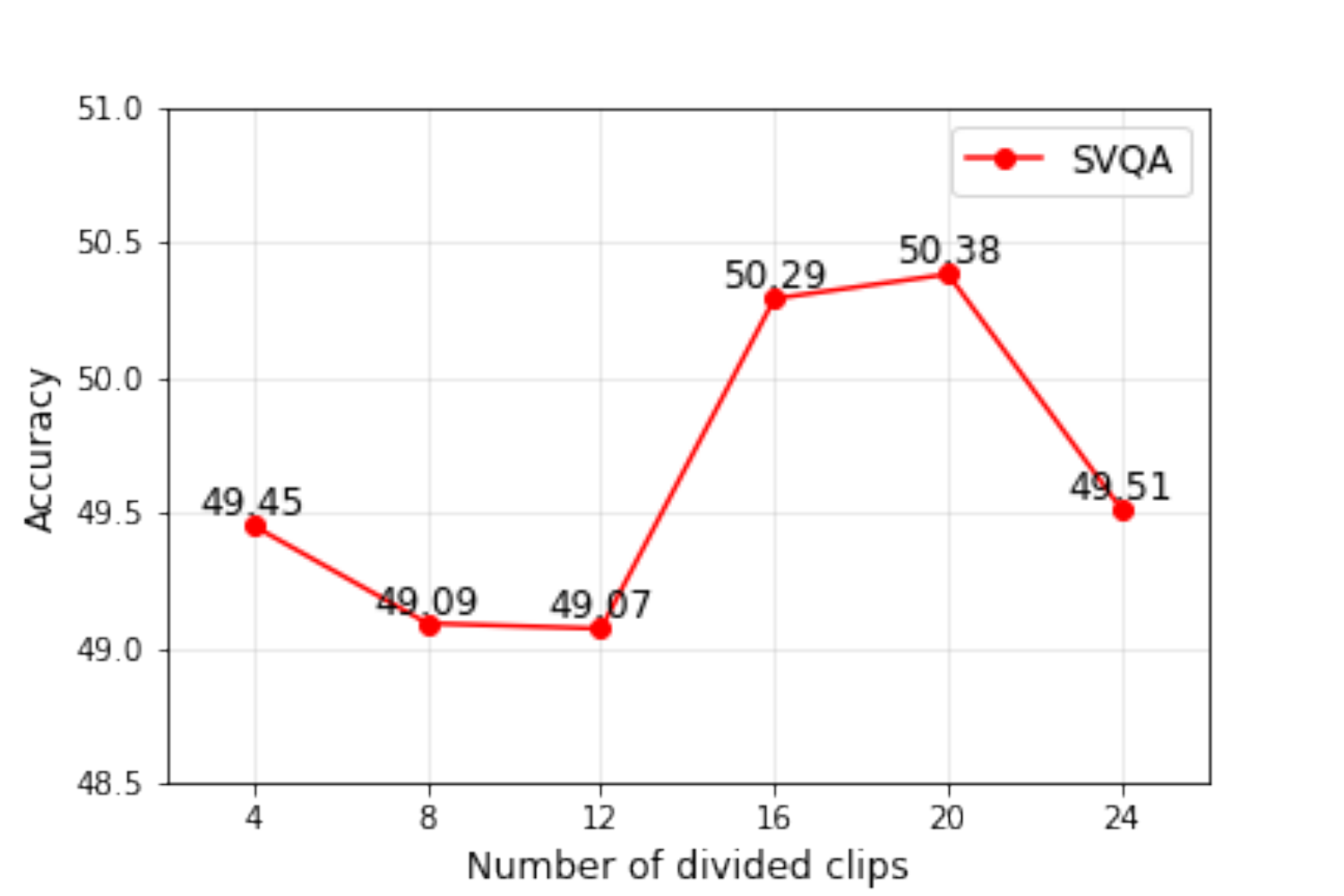}
    \caption{\textbf{Number of divided clips}. Impact of the number of divided clips in the iterative process on SVQA.}
    \label{fig:SVQA_clip}
\end{figure}
\subsubsection{The detailed analysis of the number of clips $N$} 
Since questions in SVQA are compositional, requiring agents to understand the whole video contents and spatial-temporal relations between relevant objects in videos, the key information of visual facts may distribute evenly in the videos. Therefore, we also perform a detailed analysis of the number of divided clips $N$ of SVQA dataset. We train several networks in three iteration steps on SVQA, the results are illustrated in Fig. \ref{fig:SVQA_clip}.

Networks with $16$ and $20$ clips respectively provide a gain of $+0.84\%$ and $+0.93\%$ in overall accuracy on SVQA test set over the network with $4$ clips. This implies that the performance of spatial-temporal relational modeling via Graph Neural Network could benefit from a larger number of divided clips. With more divided clips, the key information in videos would distribute more evenly, that is, we explore relational reasoning of video content in a more fine-grained manner. Finally, since each video in SVQA has the same length of $300$ frames, network with $24$ clips would contain much noise in each clip representation. Therefore, their performance degrades as the number $N$ increases.  
\subsection{Qualitative Analysis}
To better understand the contributions of our Query Punishment Module in DualVGR, we provide the visualization examples of the attention results of query features and the query-guided mask values on MSVD-QA and SVQA datasets in Fig. \ref{fig:VisualEx}.
In the first example of one step in real dataset MSVD-QA, this question intends to ask who is showing a gun in a box? Our model mainly focuses on the right part of the question ``showing a gun''. This question requires agents to focus on the motion information to infer the answer. We can observe from the example that the motion-based mask values are higher than the appearance-based mask values in all clips, which proves the effectiveness of our Query Punishment Module in real datasets. Besides, the motion-based mask values of the first two clips are higher than others. It further demonstrates the correctness of our model to pay more attention to the relevant video clips to predict the answer. For the second example of SVQA dataset, this video is extremely simple as it only contains four objects. The gray cube is rotating all the time. Then, the cyan object starts to rotate for a while. At the same time, the green cylinder begins to move upwards. In iterative step $t = 1$ , the question pays more attention to ``rotating present'', then the motion-based mask values are higher than the appearance-based mask values. In iterative step $t = 2$, the model pays much attention to ``cyan objects'', then appearance-based mask values are higher than the motion-based mask values to learn the contextual representations of cyan objects. Finally, in iterative step $t = 3$, the model considers both of them to infer the final answer. The mask values are becoming closer to each other than previous steps. Through multi-step message passing, our DualVGR progressively finds out much more question-related visual semantics to infer the answer. The visualization results further verify the effectiveness of our design.
\section{CONCLUSION}
In this paper, we propose a Dual-Visual Graph Reasoning Unit (DualVGR) for Video Question Answering, which could be stacked iteratively to model the question-related rich spatial-temporal interactions between video clips. Specifically, in our DualVGR unit, a Query Punishment Module is proposed to filter out irrelevant visual information through multiple cycles of reasoning. Then, a Video-based Multi-view Graph Attention Network is designed to learn the contextual representations of visual features to perform relational reasoning. With multi-step iterations, our DualVGR network achieves state-of-the-art or competitive performance in three mainstream VideoQA datasets, MSVD-QA, MSRVTT-QA and SVQA.

\ifCLASSOPTIONcaptionsoff
  \newpage
\fi



%

\small
\bibliography{reference}

%

\begin{IEEEbiography}[{\includegraphics[width=1in,height=1.25in,clip,keepaspectratio]{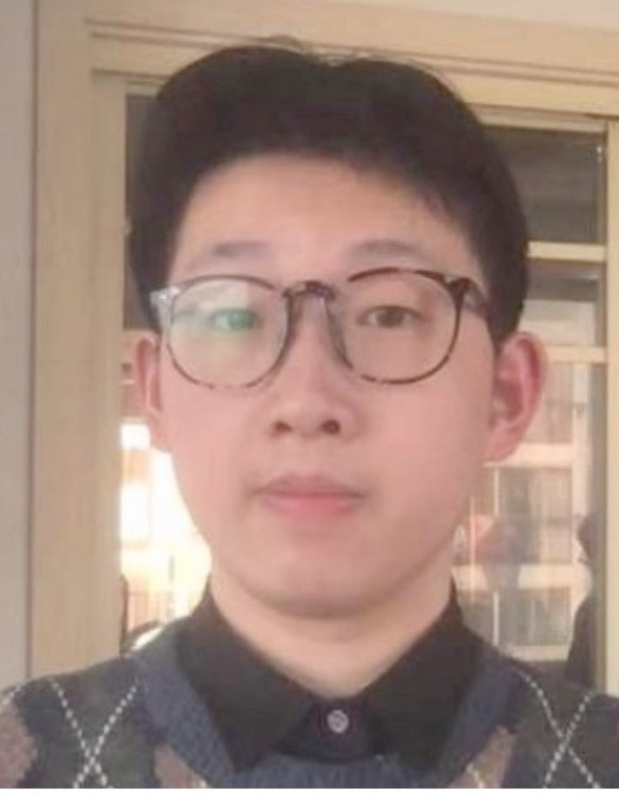}}]{Jianyu Wang}
received the bachelor's degree in electronic and information engineering from Nanjing University of Posts and Telecommunications, Nanjing, China, in 2020. He is currently pursing a Master of Computer Science Degree in University of California, San Diego, La Jolla, CA. His research interests include multimedia analysis and information retrieval, especially multimodal intelligence and recommendation system.
\end{IEEEbiography}

\begin{IEEEbiography}[{\includegraphics[width=1in,height=1.25in,clip,keepaspectratio]{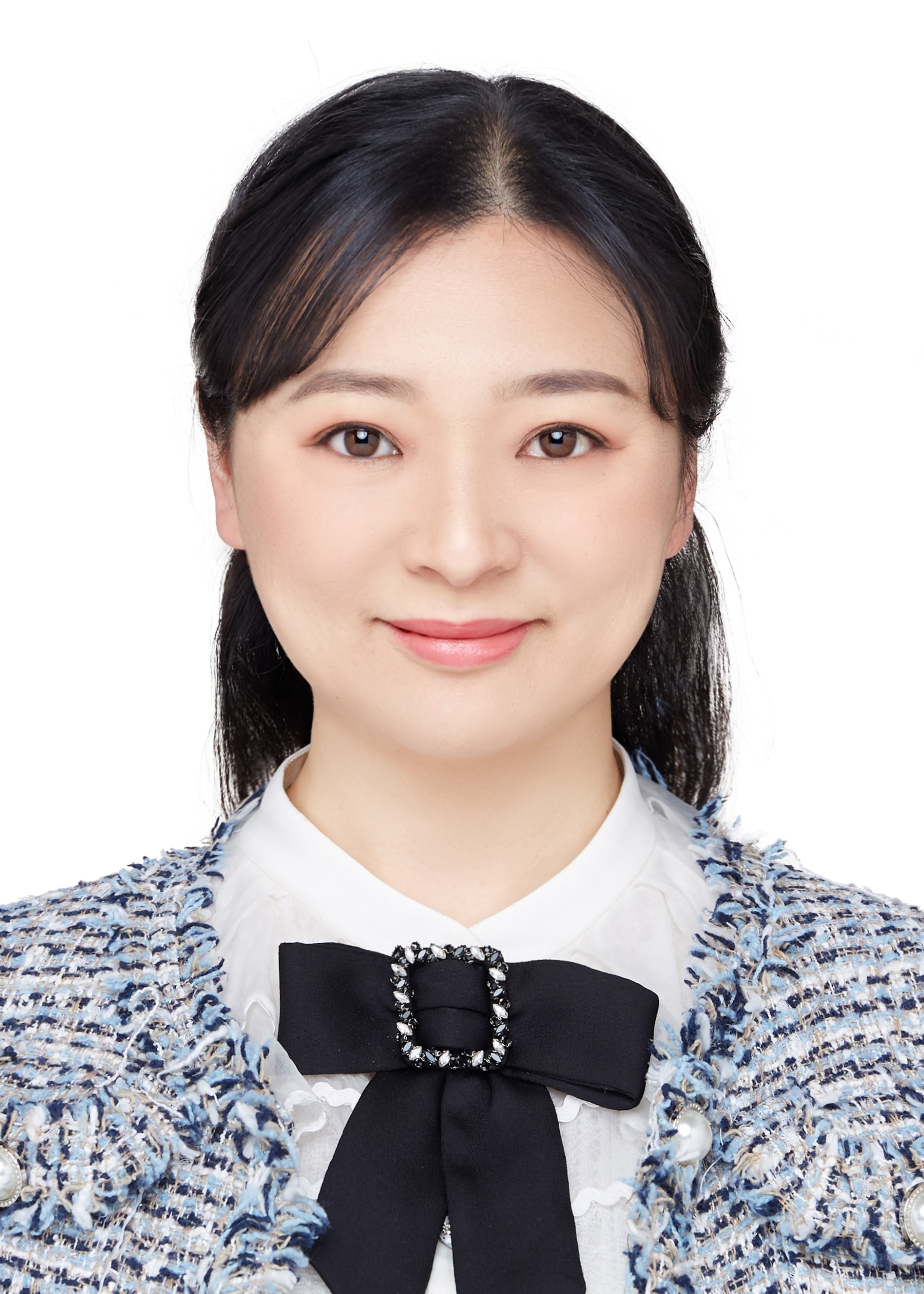}}]{Bing-Kun Bao}
is currently a professor with College of Telecommunications \& Information Engineering, Nanjing University of Posts and Telecommunications, China. Her current research interests include cross-media cross-modal image search, social event detection, image classification and annotation, and sparse/low rank representation. She was the recipient of the 2016 ACM Transactions on Multimedia Computing, Communications and Applications (ACM TOMM) Nicolas D. Georganas Best Paper Award, IEEE Multimedia 2017 Best Paper Award, the Best Paper Award from ICIMCS’09, and MMM2019 Best Paper Runner-Up Award. She is an Associate Editor of Multimedia Systems Journal, and a member of IEEE Multimedia Systems \& Applications Technical Committee. She received Outstanding Area Chair in ICME 2020.
\end{IEEEbiography}

\begin{IEEEbiography}[{\includegraphics[width=1in,height=1.25in,clip,keepaspectratio]{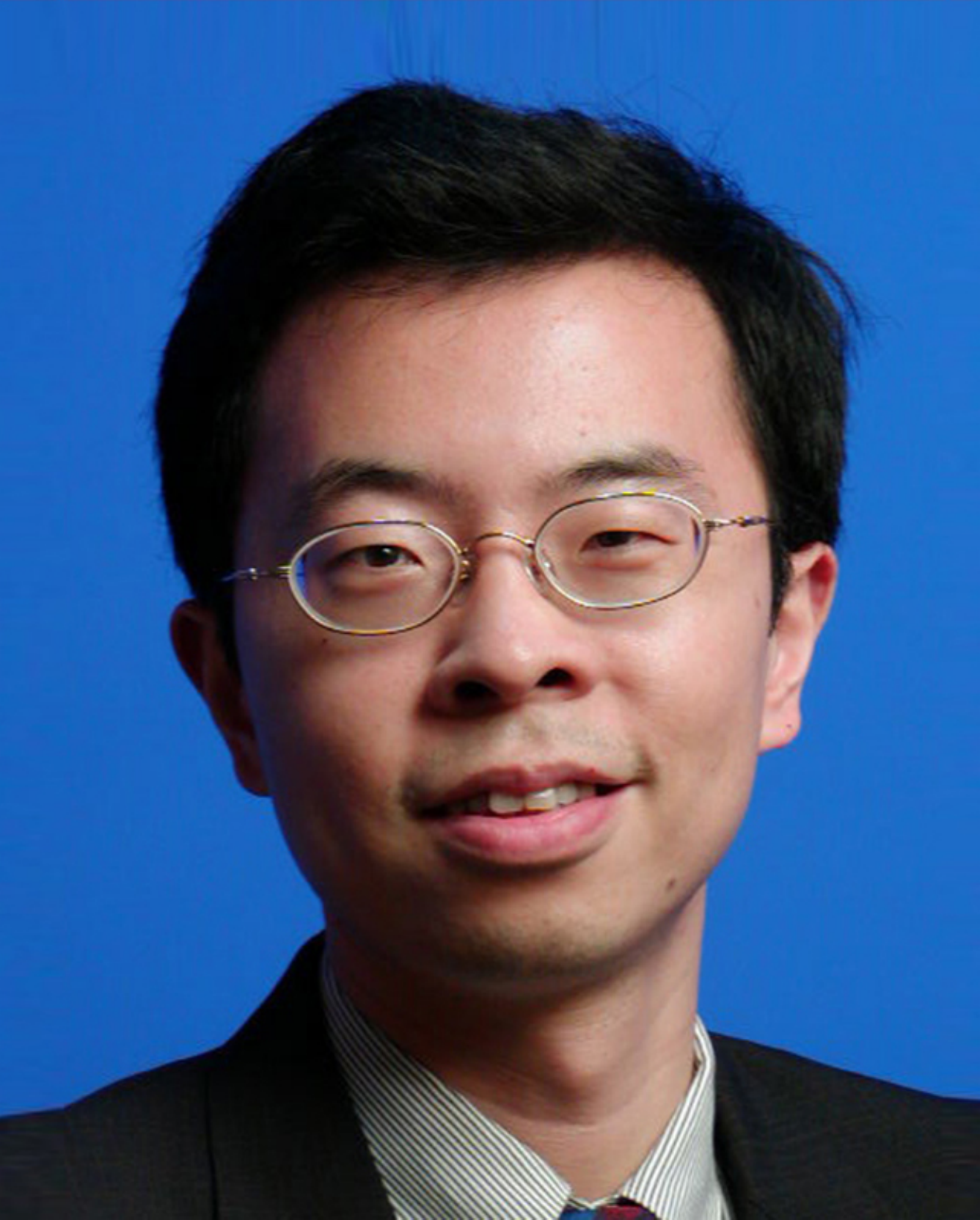}}]{Changsheng Xu}
(M’97–SM’99–F’14) is a Professor in National Lab of Pattern Recognition, Institute of Automation, Chinese Academy of Sciences. His research interests include multimedia content analysis, pattern recognition and computer vision. He has hold 50 granted/pending patents and published over 400 refereed research papers in these areas. Dr. Xu has served as editor-in-chief, associate editor, guest editor, general chair, program chair, area/track chair, and TPC member for over 20 IEEE and ACM prestigious multimedia journals, conferences and workshops, including IEEE Trans. on Multimedia, ACM Trans. on Multimedia Computing, Communications and Applications and ACM Multimedia conference. He is IEEE Fellow, IAPR Fellow and ACM Distinguished Scientist.
\end{IEEEbiography}






\end{document}